\documentclass[prb,aps,twocolumn]{revtex4-1}
\usepackage{amsmath,amssymb, graphicx}
\usepackage{bm,times, color}
\usepackage[utf8]{inputenc}
\usepackage[colorlinks]{hyperref}
\usepackage[svgnames]{xcolor}
\hypersetup{linkcolor=DarkBlue, citecolor=DarkBlue, filecolor=black, urlcolor=DarkBlue}
\usepackage{comment, mathrsfs,slashed}
\def \p{\partial}
\def \dag{\dagger}
\def \mb{\mathbf}

\def \lan{\langle}
\def \ran{\rangle}

\begin{document}
\preprint{EFI-xx}
\title{Landau quantization of multilayer graphene on a Haldane sphere}

\author{Wei-Han Hsiao}
\affiliation{Kadanoff Center for Theoretical Physics, University of Chicago,
Chicago, Illinois 60637, USA}
\date{April 2020}

\begin{abstract}
We consider the problem of multilayer graphene on a Haldane sphere and determine the Landau level spectrum for this family of systems. This serves as a generalization of the Landau quantization problem of ordinary nonrelativistic Haldane sphere and spherical graphene, or Dirac-like particles on a sphere. The Hamiltonian is diagonalized in a concise algebraic fashion exploiting two mutually commuting SU(2) algebras of the problem. Additionally, using exact wave functions we demonstrate computation of Haldane pseudopotentials in the second Landau level. These exact solutions add to the current toolkits of the numerical studies on fractional quantum Hall effects in systems of graphite multilayers. 
\end{abstract}

\maketitle

\section{Introduction}
The quantum Hall effect is one of the most fruitful subjects in modern physics. It concerns the dynamics of charged particles on a two-dimensional spatial manifold in the presence of strong magnetic field normal to the surface. It features nontrivial quantized time-reversal odd conductivity $\sigma_{xy} = -\sigma_{yx} = \nu {e^2}/{h}$, where $\nu$ can be an integer (integer quantum Hall effect) or a rational number (fractional quantum Hall effect). Despite this simple set up, ever since the discoveries of the integer and fractional quantum Hall effects \cite{PhysRevLett.45.494, PhysRevLett.48.1559}, it has been motivating vast developments in both theoretical understanding using methods such as conformal field theories (CFT) \cite{RevModPhys.89.025005}, topological quantum field theories (TQFT), field theory dualities, trial wave functions, etc. \cite{2016arXiv160606687T}, and experimental realizations in various materials including GaAs, graphene \cite{Novoselov2005}, and even synthetic materials \cite{Schine2016}. 

For both integer and fractional quantum Hall effects, the current understanding of physicists relies on an ancient problem solved by Landau. He investigated the eigenvalue problem of noninteracting nonrelativistic fermions moving on a two-dimensional slab with a uniform perpendicular magnetic field. The energy spectrum consists of flat bands with large degeneracies, which now are known as Landau levels. In addition to the original rectangular geometry considered by Landau, this Landau level problem has been posed in different gauge and geometries. In particular, occasionally it is convenient to compactify the flat slab to a sphere. The uniform magnetic field can be sourced by a Dirac or Wu-Yang \cite{WU1976365} monopole enclosed and located at the center of the sphere. The spherical model for non-interacting and non-relativistic fermions is known as the Haldane sphere \cite{PhysRevLett.51.605}. Models for Dirac-type fermions such as relativistic fermions or electrons in graphene were also addressed independently in several works \cite{JELLAL2008361, PhysRevB.93.235122, PhysRevB.94.035105, PhysRevLett.115.017001,GREITER201833}. The solutions on a sphere provide us with toolkits for numerical investigations of the fractional quantum Hall effect.

This paper aims to study the spherical model for multilayer graphene. It is an interesting problem not only because of algebraic interest. Experimentally, the integer quantum Hall effects in bilayer and trilayer graphite were discovered soon after the discovery in single-layer graphene \cite{Novoselov2005, Novoselov2006, PhysRevLett.107.126806}. Moreover, nontrivial fractional quantum Hall states were reported to exist in the zeroth, the first, and even higher Landau levels \cite{Maher61,Kou55, doi:10.1021/nl5003922, Diankov2016, Jacak2017}. Theoretical investigations have reported phase transitions in the zeroth Landau levels at various filling factor $\nu$ in bilayer graphene using a planar interaction potential \cite{PhysRevLett.112.046602} as well as the existence of potential Pfaffian states in some Landau levels \cite{PhysRevLett.107.186803}.  The explicit eigenvalues and eigenfunctions could add value to the current arsenal for numerical studies or trial wave function construction for the exotic fractional quantum Hall states in these systems. To that end, we utilize the two mutually commuting SU(2) algebras \cite{PhysRevB.83.115129} over the Hilbert space of Haldane sphere to diagonalize the models. We will show in tight binding limit, the energy spectrum for a ABC-stacked (rhombohedral) multilayer graphene is given by
\begin{align}
\label{mainResult} \varepsilon_{Qn}^J = \pm \Delta\sqrt{\prod_{k=1}^J(n-k+1)\bigg(1+\frac{n+k}{2Q}\bigg)},
\end{align}
where $\Delta$ is an energy scale dependent on the specific microscopic parametrization. $Q$, $n$, and $J$ are the charge of monopole, Landau level index, and the number of layers.

This paper is organized as follows. In Sec. \ref{formalism} we introduce the machineries required for solving the spectrum. This includes a minimal review of Haldane sphere and its two-SU(2) formulation. \cite{PhysRevLett.51.605,PhysRevB.83.115129}. In Sec. \ref{mainpart} we present the model, solve the bilayer case as a warm up, and show how the spectrum and eigenfunctions can be deduced for a general number of layers $J$. In Sec. \ref{PP} we compute the bare Haldane pseudopotentials in the second Landau level for both non-relativistic fermions and bilayer graphene as an application of the exact eigenfunctions. Then we conclude the paper. Appendix \ref{AA} contains details of the two-body matrix element of the Coulomb potential. 

\section{Formalism}\label{formalism}
\subsection{Haldane sphere and monopole harmonics}
The Haldane sphere \cite{PhysRevLett.51.605} refers to the quantum Hall problem defined on a sphere of radius $R$ enclosing a magnetic monopole. The planar momentum, promoted to a sphere, is replaced with the tangential component $\widehat{\bm{\lambda}} = \widehat{\mb r}\times(-i\nabla+e\mb A/(\hbar c))R$ with $\widehat{\mb r} = \mb R/R$. The uniform magnetic field is sourced by a Dirac monopole with magnetic flux $2Q\phi_0 = 2Q\frac{hc}{e}$. The vector potential and the magnetic field are given explicitly by
\begin{subequations}
\begin{align}
& \mb A = -\frac{\hbar cQ}{eR}\cot\theta\, \widehat{\bm{\phi}},\\
& \mb B = \frac{2Q\phi_0}{4\pi R^2}\widehat{\mb r}.
\end{align}
\end{subequations}
In this gauge, we can verify $[\widehat{\lambda}_i, \widehat{\lambda}_j] = i\epsilon_{ijk}(\widehat{\lambda}_k -Q\widehat{r}_k)$ and $[\widehat{\lambda}_i, \widehat r_j] = i\epsilon_{ijk}\widehat r_k$. This algebra motivates us to define the { angular momentum} operator 
\begin{align}
\widehat{\bm{\ell}} = \widehat{\bm{\lambda}} + Q\widehat{\mb r}.
\end{align}
We immediately see operators $\widehat{\ell}_i$ satisfy the SU(2) algebra $[\widehat{\ell}_i, \widehat{\ell}_j] = i\epsilon_{ijk}\widehat{\ell}_k$. The machinery of angular momentum can thus be utilized for computations. For example, the Hamiltonian of a non-relativistic spin-polarized fermion is $H = \frac{1}{2mR^2}\widehat{\bm{\lambda}}\cdot\widehat{\bm{\lambda}}$. Since $\widehat{\bm{\lambda}}\cdot\widehat{\mb r} = \widehat{\mb r}\cdot\widehat{\bm{\lambda}} = 0$, $\Rightarrow \widehat{\bm{\lambda}}\cdot\widehat{\bm{\lambda} }= \widehat{\bm{\ell}}\cdot\widehat{\bm{\ell}}-Q^2$. Consequently, the energy eigenvalue is $\frac{\hbar^2}{2mR^2}[\ell(\ell+1)-Q^2]$, where $\ell$ is the {orbital angular momentum quantum number} of the operator $(\widehat{\bm{\ell}})^2$. The lowest Landau level (LLL) here is indexed by the smallest value of $\min\ell = |Q|$. In the rest of the paper we focus on $Q>0$ and write $\ell = Q + n$, where $n = 0, 1,2,\dots,$ denotes the Landau level index.

Recall that in the theory of angular momentum the simultaneous eigenfunctions of $\mb L^2$ and $L_z$ form the family of spherical harmonics. Similarly, the eigenfunctions of $(\widehat{\bm\ell})^2$ and $\widehat{\ell}_z$ form another close family dubbed the monopole harmonics $Y_{Q\ell m}$\cite{WU1976365}. The quantum numbers takes value from the domains $\ell = Q, Q+1,\dots$ and $-\ell \leq m \leq \ell$. In this paper we merely include the minimal self-contained information about this special function and refer to Ref. \onlinecite{PhysRevD.16.1018} for more details. To parametrize the monopole harmonics, instead of using the polar and azimuthal angles $\theta$ and $\phi$ on a sphere, it is convenient to introduce the following spinor coordinates in the LLL:
\begin{subequations}
\begin{align}
& u = \textstyle{\cos\frac{\theta}{2}\, e^{i\phi/2}},\\
& v =\textstyle{ \sin\frac{\theta}{2}\, e^{-i\phi/2}}.
\end{align}
\end{subequations}
In particular, $Y_{QQm}\sim u^{Q+m}v^{Q-m}$. They can be regarded as the spherical counterparts of the holomorphic coordinate $z$ on a disk. The orbital angular momentum operators can be written as differential operators of $u$ and $v$ in the LLL. Explicitly, $\widehat{\ell}_z = \frac{1}{2}(u\p_u-v\p_v)$, $\widehat{\ell}_+ = u\p_v$, and $\widehat{\ell}_- = v\p_u$. One can straightforwardly verify $\widehat{\ell}_zY_{QQm} = mY_{QQm}$ and $(\widehat{\bm{\ell}})^2Y_{QQm} = Q(Q+1)Y_{QQm}.$ In the next section we discuss the generalization to higher Landau levels and introduce the other set of SU(2) operators.

\subsection{Two-SU(2) formulation}
As we just mentioned, analogous with the fact that wave functions in the lowest Landau level physics on a disk are holomorphic, the wave functions in the lowest Landau level on a Haldane sphere consist of only spinors $u$ and $v$. The conjugate spinors $\bar u$ and $\bar v$ come into play as one wishes to incorporate higher Landau levels. The complete expression of $\widehat{\bm{\ell}}$ was constructed by Greiter \cite{PhysRevB.83.115129} as below
\begin{subequations}
\begin{align}
& \widehat{\ell}_+ = \widehat{\ell}_x + i\widehat{\ell}_y = u\frac{\p}{\p v} - \bar v\frac{\p}{\p\bar u},\\
& \widehat{\ell}_- = (\widehat{\ell}_+)^{\dag} = v\frac{\p}{\p u} - \bar u\frac{\p}{\p\bar v},\\
& \widehat{\ell}_z = \frac{1}{2}\bigg(u\frac{\p}{\p u} - v\frac{\p}{\p v}-\bar u\frac{\p}{\p\bar u} +\bar v\frac{\p}{\p\bar v}\bigg).
\end{align}
\end{subequations}
Moreover, Greiter discovered another set of operators $\widehat{s}_i$ defined as
\begin{subequations}
\begin{align}
& \widehat{s}_+ = \widehat{s}_x + i\widehat{s}_y = u\frac{\p}{\p\bar v} - v\frac{\p}{\p\bar u},\\
& \widehat{s}_- = (\widehat s_+)^{\dag} = \bar v\frac{\p}{\p u} -\bar u \frac{\p}{\p v},\\
& \widehat s_z = \frac{1}{2}\bigg(u\frac{\p}{\p u} + v\frac{\p}{\p v} -\bar u\frac{\p}{\p\bar u} -\bar v\frac{\p}{\p\bar v}\bigg).
\end{align}
\end{subequations}
It is straightforward to show
\begin{align}
\label{sSU(2)}& [\widehat s_i, \widehat s_j] = i\epsilon_{ijk}\widehat s_k,
\end{align}
and $[\widehat{\ell}_i,\widehat{s}_j] = 0$. We will see shortly the Hilbert space of the Haldane sphere problem is described by these two mutually commuting algebras, which is analogous to the pair of mutually commuting ladder operators in the Landau quantization problem on a disk.

Given the second set of SU(2) algebra, we also would like to construct proper eigenstates of operators $\widehat{\mb s}^2$ and $\widehat{s}_z$. Owing to the fact these two algebras commute, the eigenfunctions are members of monopole harmonics and we simply have to identify the eigenvalues. Let us first observe $\widehat{\mb r} = (\bar uv + \bar vu, i(\bar uv-\bar vu), \bar uu-\bar vv)$ and hence $\widehat{\mb r}\cdot\widehat{\bm{\ell}} = \widehat{s}_z$, implying 
\begin{align}
\widehat{s}_zY_{Q\ell m} = (\widehat{\mb r}\cdot\widehat{\bm{\ell}}\, )Y_{Q\ell m} = QY_{Q\ell m}.
\end{align}
Next, we further notice that $\frac{1}{2}[\widehat{\ell}_+\widehat{\ell}_- + \widehat{\ell}_-\widehat{\ell}_+ -\widehat s_+\widehat s_--\widehat s_-\widehat s_+] = (\widehat s_z)^2-(\widehat{\ell}_z)^2$ and hence $(\widehat{\bm{\ell}})^2 = \widehat{\mb s}^2$. As a consequence, 
\begin{align}
\widehat{\mb s}^2Y_{Q\ell m} = \ell(\ell+1)Y_{Q\ell m}.
\end{align}
We have by far confirmed the monopole harmonics $Y_{Q\ell m}$ is a common eigenstate of both $((\widehat{\bm{\ell}})^2,\widehat{\ell}_z)$ and $(\widehat{\mb s}^2, \widehat s_z)$. The major difference is the action of ladder operators $\widehat{\ell}_{\pm}$ and $\widehat s_{\pm}$. The former creates or annihilates $m$ by one unit, keeping $Q$ and $\ell$ fixed. On the other hand, the latter fixes $m$ and $\ell$ while creating or annihilating $Q$ by one unit. Since $\ell = Q + n$, decreasing $Q$ is equivalent to raising the Landau level index if $\ell$ is fixed. To make this explicit, for action of $\widehat{s}_i$ we use a new notation
\begin{align}
\mathscr Y_{Qnm} := Y_{Q\ell m}
\end{align}
and the actions of $\widehat{s}_i$ read 
\begin{subequations}
\begin{align}
\label{sz}& \widehat s_z\mathscr Y_{Qnm} = Q\mathscr Y_{Qnm},\\
\label{sp}& \widehat{s}_+\mathscr Y_{Qnm} = \sqrt{n(1+n+2Q)}\, \mathscr Y_{(Q+1)(n-1)m},\\
\label{sm}& \widehat{s}_-\mathscr Y_{Qnm} = \sqrt{(n+1)(n+2Q)}\, \mathscr Y_{(Q-1)(n+1)m}.
\end{align}
\end{subequations}
The action $\widehat{s}_-\widehat s_+$ effectively translate to the multiplication $n(n+1)+2nQ$. Equations~\eqref{sz},~\eqref{sp}, and~\eqref{sm} are the main tools we will utilize in the following sections.
\section{Multilayer graphene on a Haldane sphere}\label{mainpart}
We consider the following family of Hamiltonians parametrized by $J=1,2,\dots$:
\begin{align}
\label{flat} H = \varepsilon_c\begin{pmatrix} 0 & (\pi^{\dag})^J\\ \pi^J & 0\end{pmatrix},
\end{align}
where $\pi= (-i\p_x+A_x)-i(-i\p_y+A_y)$. For $J = 1$, it reduces to the well known Dirac Hamiltonian or graphene. It can be shown that in the tight binding limit, the low-energy Hamiltonian of the AB-stacked bilayer graphene and ABC-stacked (rhombohedral) trilayer graphene on a plane assume this form with $J =2$ and $3$\cite{McCann_2013}. $\varepsilon_c$ is the parameter specifying the energy scale. It can be the Fermi velocity $v_F$ for $J=1$ and is conventionally the band curvature or inverse mass $-\frac{1}{2m}$ for $J=2$. In the following we will first review basic facts of the planar model for $J=2$ and then solve its spherical generalization. From there we proceed to tackle models on the Haldane sphere for any positive integer $J$.
\subsection{$J=2$}
Taking $J= 2$ and $\varepsilon_c = -\frac{1}{2m}$, the energy spectrum of~\eqref{flat} in a perpendicular magnetic field is given by $\pm \omega_c\sqrt{n(n-1)}$\cite{PhysRevLett.96.086805}. In particular, the zero-energy bands are determined by the condition $(\pi^{\dag})^2\psi = 0$ and thus there exist two bands at exactly zero energy. The immediate peculiarity is that non-holomorphic functions $\bar zz^me^{-\frac{1}{4}|z|^2}$ come into play even in the lowest Landau level because of this signature of multi-zero-energy bands. 

To place Eq.~\eqref{flat} on to a sphere of radius $R$, as shown in Ref. \onlinecite{PhysRevB.94.035105}, one ought to promote $\pi$ to the components tangent to the sphere $R^{-1}(\widehat{\lambda}_{\theta} - i\widehat{\lambda}_{\phi})= R^{-1}\widehat s_+$, where $\widehat{\lambda}_{\theta}$ and $\widehat{\lambda}_{\phi}$ are the components of $\widehat{\bm\lambda}$ in the directions of $\hat{\bm{\theta}}$ and $\hat{\bm{\phi}}$, respectively. Therefore, for $J = 2$,
\begin{align}
\label{spherical}H_{b} = -\frac{1}{2mR^2}\begin{pmatrix} 0 & (\widehat{s}_+)^2 \\ (\widehat{s}_-)^2 & 0\end{pmatrix}.
\end{align}
The form of the Hamiltonian is the main reason we adapt two-SU(2) formulation since $\widehat s_i$'s are more natural variables than $\widehat{\ell}_i$'s for these computations. To solve Eq.~\eqref{spherical}, we first square it to get
\begin{align}
H_b^2 =\frac{1}{4m^2R^2} \begin{pmatrix} (\widehat{s}_+)^2(\widehat{s}_-)^2 & 0 \\  0 & (\widehat{s}_-)^2(\widehat{s}_+)^2\end{pmatrix}.
\end{align}
Using the SU(2) algebras~\eqref{sSU(2)},
\begin{subequations} 
\begin{align}
\label{11}(\widehat{s}_+)^2(\widehat{s}_-)^2 = &(\widehat{s}_-\widehat{s}_+)^2+6\widehat{s}_z(\widehat{s}_-\widehat{s}_+)-2\widehat{s}_-\widehat{s}_+ \notag\\
&+8\widehat{s}_z^2 -4\widehat{s}_z\\
\label{22}(\widehat{s}_-)^2(\widehat{s}_+)^2 =& (\widehat{s}_-\widehat{s}_+)^2- 2\widehat{s}_z(\widehat{s}_-\widehat{s}_+) -2\widehat{s}_-\widehat{s}_+
\end{align}
\end{subequations}
As we showed in the previous section, the operators on the diagonal can be diagonalized by the monopole harmonics $\mathscr Y_{Qnm}$ and thus the eigenstates of $H_b^2$ takes the form $(\mathscr Y_{Q'n'm'}, \mathscr Y_{Qnm})^T$. In order for $(\widehat{s}_+)^2(\widehat{s}_-)^2\mathscr Y_{Q'n'm'}$ and $(\widehat{s}_-)^2(\widehat{s}_+)^2\mathscr Y_{Qnm}$ to give the same energy eigenvalue, we find $Q' = Q+2$ and $n' = n-2$, and therefore for $n>1$ the state $(\mathscr Y_{(Q+2)(n-2)m}, \mathscr Y_{Qnm})^T$
\begin{comment}
\begin{align}
\label{eigenfunction} |Qnm ) = \frac{1}{\sqrt{2}}\begin{pmatrix} 
\mathscr Y_{(Q+2)(n-2)m}\\
\mathscr Y_{Qnm}
\end{pmatrix}
\end{align}
\end{comment}
is an eigenvector of $(H_b)^2$ associated with eigenvalue 
\begin{align}
\label{evalSq}\omega_c^2\bigg[n(n-1)\bigg(1+\frac{n+1}{2Q}\bigg)\bigg(1+\frac{n+2}{2Q}\bigg)\bigg],
\end{align}
where we have identified $\frac{|Q|}{mR^2}$ to be the cyclotron frequency $\omega_c$. 
From  Eq.~\eqref{evalSq} we infer the spectrum of model~\eqref{spherical} to be
\begin{align}
\label{bilayer} \pm \omega_c\sqrt{\bigg[n(n-1)\bigg(1+\frac{n+1}{2Q}\bigg)\bigg(1+\frac{n+2}{2Q}\bigg)\bigg]}.
\end{align}
As a consistency check, we look at the planar limit $Q\to \infty$. $E_{\infty nm}\to \pm \omega_c\sqrt{n(n-1)}.$ The bands $n = 0$ and $n=1$ correspond to the zero modes of the model. In these cases, the eigenstate only has the lower entry $|Qnm\ran = (0, \mathscr Y_{Qnm})^T$. Equation~\eqref{bilayer} can be written in a more compact form by extending the domain of $n$ from $\mathbb Z^+$ to $\mathbb Z$. For $|n|>2$, we replace the sign of the band energy $\pm$ and band index $n$ with sgn$(n)$ and $|n|$ and the eigenstates read
\begin{align}
\label{eigenfunction} |Qnm ) = \frac{1}{\sqrt{2}}\begin{pmatrix} 
\mathrm{sgn}(n)\mathscr Y_{(Q+2)(|n|-2)m}\\
\mathscr Y_{Q|n|m}
\end{pmatrix}.
\end{align}
%\begin{figure}
  %\includegraphics[width=\linewidth]{figure1}
  %\caption{Plot of mass corrected by Landau parameter $F_2$}
  %\label{mass}
%\end{f}
\subsection{$J>2$}
Comparing Eq.~\eqref{bilayer} and the single-layer graphene result, we can observe a pattern a propose a generalization to multilayer graphene of $J$ layers in the tight binding limit. Algebraically, it amounts to diagonalizing the Hamiltonian 
\begin{align}
H_J^2 = \begin{pmatrix} (\widehat{s}_+)^J(\widehat{s}_-)^J & 0 \\ 0 & (\widehat{s}_-)^J(\widehat{s}_+)^J\end{pmatrix}.
\end{align}
By inspection, for this model we consider the ansatz $|\psi^J_{Qn}) = (\mathscr Y_{(Q+J)(n-J)}, \mathscr Y_{Qn})^T$ associated with the eigenvalue 
\begin{align}
\label{conjEval}\prod_{k=1}^{J}(n-k+1)(n+k+2Q).
\end{align}
The quantum number $m$ is suppressed here for notational simplicity. The ansatz is consistent with the known results for $J = 1$ \cite{PhysRevB.93.235122, PhysRevB.94.035105, GREITER201833} and $J = 2$. We would like to show the it persists for general $J$. We first look at $(\widehat{s}_-)^{J}(\widehat{s}_+)^{J}\mathscr Y_{Qn}$. Repeatedly applying Eqs.~\eqref{sp} and~\eqref{sm},
\begin{widetext}
\begin{subequations}
\begin{align}
& (\widehat{s}_-)^{J}(\widehat{s}_+)^{J}\mathscr Y_{Qn} = (s_-)^J\prod_{k=1}^{J}\sqrt{(n-k+1)(k+n+2Q)}\, \mathscr Y_{(Q+J)(n-J)}\notag\\
= &  \prod_{k=1}^{J}\sqrt{(n-k+1)(n+k+2Q)}\prod_{k=J}^1\sqrt{(n-k+1)(n+k+2Q)}\, \mathscr Y_{Qn} =\prod_{k=1}^{J}{(n-k+1)(n+k+2Q)} \mathscr Y_{Qn}.
\end{align}
By exactly the same token, the term $(\widehat{s}_+)^{J}(\widehat s_-)^{J}\mathscr Y_{(Q+J)(n-J)}$ can be evaluated as
\begin{align}
&(\widehat s_+)^J(\widehat s_-)^J\mathscr Y_{(Q+J)(n-J)} = \prod^{J}_{k=1}(n-k+1)(n+k+2Q)\, \mathscr Y_{(Q+J)(n-J)}.
\end{align}
\end{subequations}
\end{widetext}
\begin{comment}
\begin{widetext}
\begin{subequations}
\begin{align}
& (\widehat{s}_-)^{J+1}(\widehat{s}_+)^{J+1}\mathscr Y_{Qn} = \widehat{s}_-(\widehat{s}_-)^J(\widehat{s}_+)^J\widehat{s}_+\mathscr Y_{Qn} = \widehat{s}_-(\widehat{s}_-)^J(\widehat{s}_+)^J\sqrt{n(1+n+2Q)}\, \mathscr Y_{(Q+1)(n-1)}\notag\\
= & \widehat{s}_-\prod_{k=1}^J(n-k)(1+n+k+2Q)\sqrt{n(1+n+2Q)}\, \mathscr Y_{(Q+1)(n-1)} = \prod_{k=1}^{J+1}(n-k+1)(n+k+2Q)\, \mathscr Y_{Qn}.
\end{align}
Next we tackle the more complex term $(\widehat{s}_+)^{J+1}(\widehat s_-)^{J+1}\mathscr Y_{(Q+J+1)(n-J-1)}.$
\begin{align}
& \widehat s_+(\widehat s_+)^J(\widehat s_-)^J\widehat s_-\mathscr Y_{(Q+J+1)(n-J-1)} = \sqrt{(n-J)(1+J+n+2Q)}\widehat s_+(\widehat s_+)^J(\widehat s_-)^J\mathscr Y_{(Q+J)(n-J)}\notag\\
= & \sqrt{(n-J)(1+J+n+2Q)}\, \widehat s_+\prod_{k=1}^J(n-k+1)(n+k+2Q)\mathscr Y_{(Q+J)(n-J)} \notag\\
=& \prod^{J+1}_{k=1}(n-k+1)(n+k+2Q)\, \mathscr Y_{(Q+J+1)(n-J-1)}.
\end{align}
\end{subequations}
\end{widetext}
\end{comment}
The conjecture~\eqref{conjEval} is then confirmed. Taking the square root of Eq.~\eqref{conjEval} and extracting from it the factor of $(2Q)^{J/2}$, we arrive at Eq.~\eqref{mainResult} and complete the derivation. 

This spectrum has the following feature. The Landau levels indexed from 0 to $J-1$ all collapse to a degenerate zero-energy band, each has degeneracy $2(Q+n)+1$. The first excited band indexed $n = J$ has a gap proportional to the numerical factor $\sqrt{J!}$. Consequently, for a reasonably large $J$, we have abundant low-energy states and a giant gap separating them from the first excited state.  As a side product, from Eq.~\eqref{mainResult}, we can also show the spectrum of model~\eqref{flat} approaches $\varepsilon_c\sqrt{n(n-1)\cdots (n-J+1)}$ by taking $Q\to\infty$, which can be able be shown by using ladder operators in disk geometry. Algebraically, it comes from the action $a^J|n\ran = [{n(n-1)\cdots(n-J+1)}]^{1/2}\, |n-J\ran$, where $a$ is a lowering operator in a simple harmonic oscillator problem and $|n\ran$ is the eigenstate. 

For $n = 0,1,\dots, J-1$, the eigenstate has only the lower entry $|Qnm) =(0,\mathscr Y_{Qnm})^T $. The rest of the eigenstates can again be parametrized compactly by promoting $n$ from an unsigned integer to a signed one. For $|n|>J-1$,
\begin{align}
|\psi^J_{Qnm}) = \frac{1}{\sqrt{2}} \begin{pmatrix} \mathrm{sgn}(n)\mathscr Y_{(Q+J)(|n|-J)m}\\ \mathscr Y_{Q|n|m}\end{pmatrix}.
\end{align}
\section{Bare Pseudopotential}\label{PP}
A direct application of the eigenstates here is the computation of the Haldane pseudopotential $V_{L}$ for a quantum Hall system on a sphere. It represents the energy of a pair of particles with a specified relative angular momentum $L$. In the fractional quantum Hall regime, the kinetic energy is quenched within each Landau level and the pseudopotential effectively determines the model Hamiltonian. Let us consider a two-body Coulomb interaction between particle 1 and particle 2 and its parametrization in term of pseudopotential.
\begin{align}
V(\mb r_1,\mb r_2) =\frac{e^2}{\epsilon} \frac{1}{|\mb r_1- \mb r_2|} =\sum_{L}V_{L}\widehat{P}_{12}(L).
\end{align}
$\widehat{P}_{12}(L)$ is the projector of the subspace where the relative angular momentum between particle 1 and 2 is $L$. On a sphere, the distance $|\mb r_1-\mb r_2|$ is often replaced with the chord distance ${2}\, R|u_1v_2 - u_2v_1|$. $V_{L}$ can be computed by inverting the above 
\begin{align}
V_{L} = \sum_{\{ m_i\}}&\bigg[ \lan L m  |\ell_1m_1'\ell_2m_2'\ran \lan \ell_1m_1\ell_2m_2|L m\ran\notag\\
&(1',2'|V|1,2)\delta_{L,m_1+m_2}\delta_{m_1'+m_2',m_1+m_2}\bigg],
\end{align}
where we use the notation $|\ )$ to denote the doublet states. Here $|1,2) = |\ell_1m_1)\otimes |\ell_2m_2)$ is the tensor product state of two particles labeled by the quantum number $(\ell_i,m_i)$. $\ell_i =|Q|+n $ is the shell angular momentum for particle $i$. $ \lan \ell_1m_1\ell_2m_2|L m\ran$ is the Clebsch-Gordan coefficients projecting the two-particle state to the state with total angular momentum $(L,m)$. Each angular momentum in the sum $m_i$ ranges over $-\ell_i \leq m_i \leq \ell_i$. Non-vanishing summands must satisfy $m_1+m_2 = m_1' + m_2'$ and $\ell_1 + \ell_2 \leq L$.

Let us demonstrate how to utilize previous results in a sample computation for $J=2$, the model of bilayer graphene. In this case, there are two flat bands at the zeroth and the first Landau levels. The corresponding eigenvectors have only the lower entry and thus the bare pseudopotentials are identical to those of the non-relativistic fermions. Here we would like to focus on the first excited state in the second Landau level $n=2$. Given $Q$ and $n$, with~\eqref{eigenfunction} we can write
\begin{widetext}
\begin{align}
\label{ELE1} (1',2'|V|1,2) =& \frac{1}{4}[\lan(Q+2)m_1'(Q+2)m_2'|V|(Q+2)m_1(Q+2)m_2\ran +\lan Qm_1'Qm_2'|V|Qm_1Qm_2\ran\notag\\
&+ \lan (Q+2)m'_1Qm'_2|V|(Q+2)m_1Qm_2\ran + \lan Qm_1'(Q+2)m_2'|V|Qm_1(Q+2)m_2\ran ],
\end{align}
\end{widetext}
where $\lan Q_1m_1'Q_2m_2'|V|Q_1m_1Q_2m_2\ran$ is the two-body matrix element for a pair of non-relativistic fermions . We present the derivation and explicit form of this matrix in terms of integrals of monopole harmonics in Appendix \ref{AA}. The result is plotted in Fig.~\ref{pseudoP} for a monopole $2Q =7$. The same quantities of the non-relativistic fermions $V_{L}^{\rm nr}$ in the second and the lowest Landau level are shown for the sake of comparisons. Note that instead of $V_{L}$, we plot $V_{2L-m}$, which in the planar limit $R\to\infty$ approaches the planar pseudopotential of a pair of particles with relative angular momentum $m$. The trend in Fig.\ref{pseudoP} is reminiscent of the comparison of the first Landau levels of graphene and that of non-relativistic fermions \cite{PhysRevB.93.235122}. In the $n$th Landau level where $n>0$, $V_{2L-m}^{\rm nr}$ typically has oscillatory feature at small $m$ and eventually becomes monotonically decreasing as $m$ approaches $2L$. Such oscillation renders the stability of the Laughlin state. On the other hand, in the second Landau level of the bilayer graphene, these bumps are smeared and the bare pseudopotential is monotonic in $m$. These tiny bumps are not finite-size artifacts. Though they are flattened by increasing $Q$, they persist in the planar limit \cite{ajit_2020}. Our result suggests a relatively stronger repulsion at small $m$ in the second Landau level of the bilayer graphene compared to the non-relativistic fermions. The genuine nature of the many-body ground state cannot be told without more sophisticated investigations.
\begin{figure}
 \includegraphics[width=0.9\linewidth]{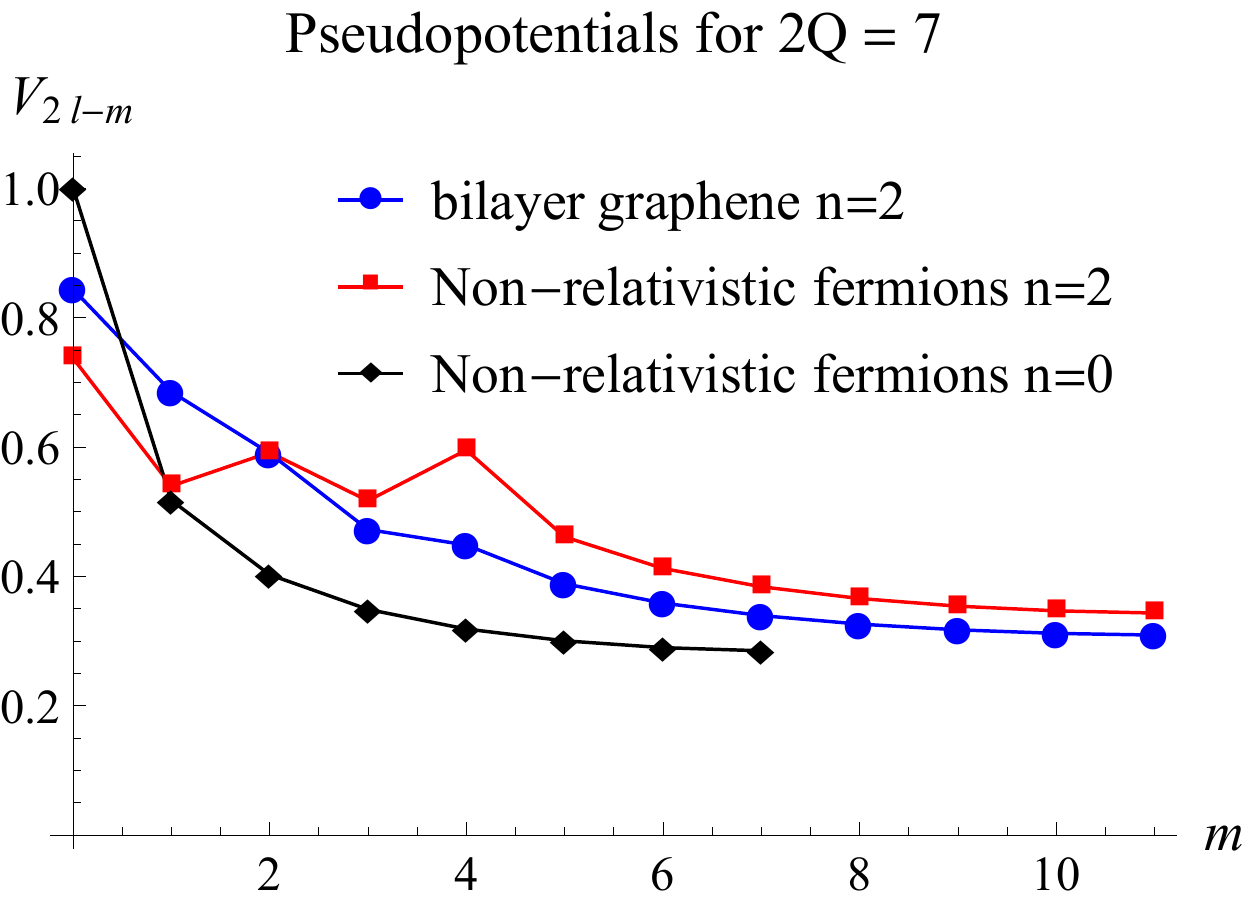}
 \caption{Pseudopotentials in the unit of $e^2/(\epsilon\ell_B)$ on a Haldane sphere of $2Q = 7$ for non-relativistic fermions with Landau level index $n=0, 2$ and the bilayer graphene system with $n=2$.}
 \label{pseudoP}
\end{figure}

We leave the thorough numerical studies for another separate work and close this section with a comment. Bare pseudopotentials receive corrections from Landau level mixing, the strength of which is characterized by the ratio of Coulomb energy to cyclotron frequency $\kappa$. A systematic approach for deriving these corrections perturbatively in $\kappa$ was introduced in Ref. \onlinecite{PhysRevB.87.245129}. For the first $J$ Landau levels in a multilayer graphene the perturbative approach seems to fail because of the degeneracy at zero energy and requires a scalable non-perturbative alternative. 

\section{Conclusion}
We studied a family of quantum mechanics models~\eqref{flat} defined on the Haldane sphere. They provide low-energy descriptions for multilayer graphene of number of layers greater than one. The spectra and associated eigenfunctions were solved using the two-SU(2) formalism. The exact eigenfunctions are pivotal ingredients for numerical studies of fractional quantum Hall phases of matter and computations of Haldane pseudopotentials. We also computed the bare pseudopotentials for the bilayer graphene from a Coulomb interaction and showed the potential profile monotonically decreases in relative angular momentum $2L-m$, which deviates from the non-relativistic counterpart qualitatively. Various questions can be posed based on the present work. In particular, since multiple Landau levels become degenerate at zero energy, one naturally could pursue the generalization of Laughlin states consisting of single particle wave functions from multiple Landau levels. Together with recent discoveries of quantum Hall phenomena in graphite multilayers, the results here are anticipated to motivate future works on strongly correlated electronic physics in the realm of fractional quantum Hall effect.

\acknowledgements
The author thanks Ajit Balram for his valuable comments on the manuscript and sharing unpublished information. This work is supported, in part, by U.S.\ DOE Grant
No.\ DE-FG02-13ER41958 and a Simons Investigator Grant from the Simons Foundation.  

\appendix
\section{The matrix elements}\label{AA}
In this Appendix we write down the closed-form result for the matrix element $\lan Q_1m_1'Q_2m_2'|V|Q_1m_1Q_2m_2\ran$ involved in Eq.~\eqref{ELE1} in terms of integrals of monopole harmonics and evaluate them using the addition theorems. We follow the conventions in Ref. \onlinecite{WU1976365}. A more modern introduction can be found in Ref. \onlinecite{jain_2007}. First, we expend the Coulomb potential in terms of the spherical harmonics and identify $Y_{\ell}^m = Y_{(Q=0)\ell m}$. The two-body matrix element then assumes the following form:
\begin{widetext}
\begin{align}
& \lan Q_1 m_1' Q_2m_2'|V|Q_1m_1Q_2m_2\ran =  \frac{e^2}{\epsilon}\int d\Omega_1d\Omega_2 Y^*_{Q_1\ell m_1'}(\mb r_1)Y^*_{Q_2\ell m_2}(\mb r_2) \frac{1}{|\mb r_1-\mb r_2|}Y_{Q_1\ell m_1}(\mb r_1)Y_{Q_2\ell m_2}(\mb r_2)\notag\\
\label{MELE}= &\frac{4\pi e^2}{\epsilon R}\sum_{\ell'=0}^{\infty}\sum_{m' = -\ell'}^{\ell'}\frac{1}{2\ell'+1}\int d\Omega_1d\Omega_2\, Y^*_{Q_1\ell m_1'}(\mb r_1)Y^*_{0\ell'm'}(\mb r_1)Y_{Q_1\ell m_1}(\mb r_1)Y^*_{Q_2\ell m_2'}(\mb r_2)Y_{0\ell'm'}(\mb r_2)Y_{Q_2\ell m_2}(\mb r_2).
\end{align}
Using the addition theorems for monopole harmonics, the integrals have closed form solutions as follows:
\begin{align}
&\int d\Omega_1\, Y^*_{Q_1\ell m_1'}(\mb r_1)Y^*_{0\ell'm'}(\mb r_1)Y_{Q_1\ell m_1}(\mb r_1)\notag\\
= & (-1)^{Q_1-m_1'-m'+\ell'}\sqrt{\frac{(2\ell+1)^2(2\ell'+1)}{4\pi}}\begin{pmatrix} \ell & \ell' & \ell \\ m_1' & m' & -m_1\end{pmatrix}\begin{pmatrix} \ell & \ell' & \ell \\ -Q_1 & 0 & Q_1\end{pmatrix}.
\end{align}
\begin{align}
& \int d\Omega_2\, Y^*_{Q_2\ell m_2'}(\mb r_2)Y_{0\ell'm'}(\mb r_2)Y_{Q_2\ell m_2}(\mb r_2)\notag\\
= & (-1)^{Q_2-m_2'+\ell'}\sqrt{\frac{(2\ell+1)^2(2\ell'+1)}{4\pi}}\begin{pmatrix} \ell & \ell' & \ell \\ m_2' & -m' & -m_2\end{pmatrix}\begin{pmatrix} \ell & \ell' & \ell \\ -Q_2 & 0 & Q_2\end{pmatrix}.
\end{align}
The big parenthesis denotes Wigner-$3j$ symbol. Triangle inequality satisfied by $3j$ symbol terminates the infinite sum at $\ell_{\mathrm{max}}'=2\ell$. Combing these integrals together, we obtain 
\begin{align}
&\lan Q_1 m_1' Q_2m_2'|V|Q_1m_1 Q_2m_2\ran /[\frac{e^2}{\epsilon R}(2\ell+1)^2] \notag\\
\label{finalform}= & \sum_{\ell'=0}^{2\ell}\sum_{m=-\ell'}^{\ell'}(-1)^{Q_1+Q_2-m_1'-m_2'-m'}\begin{pmatrix} \ell & \ell' & \ell \\ m_1' & m' & -m_1\end{pmatrix}\begin{pmatrix} \ell & \ell' & \ell \\ -Q_1 & 0 & Q_1\end{pmatrix}\begin{pmatrix} \ell & \ell' & \ell \\ m_2' & -m' & -m_2\end{pmatrix}\begin{pmatrix} \ell & \ell' & \ell \\ -Q_2 & 0 & Q_2\end{pmatrix}.
\end{align} 
\end{widetext}
Equation~\eqref{finalform} can be computed using package programs such as {\it Mathematica} and SYMPY. Finally we have to express $R$ in terms of physical measure. In a quantum Hall problem, the length scale is set by the magnetic length $\ell_B = \sqrt{{\hbar c}/{(eB)}}$. The magnitude of the magnetic field produced by the monopole is $B = \hbar c |Q|/(eR^2)$. Hence we replace $R$ with $R = \sqrt{|Q|}\sqrt{\hbar c/(eB)} = \sqrt{Q}\, \ell_B$ and the pseudopotentials computed are presented in the unit of $e^2/(\epsilon\ell_B)$. Consequently, given a fixed $|B|$, the planar limit $R\to\infty$ is then equivalent to $Q\to\infty$.

\bibliography{citation}{}
\bibliographystyle{apsrev4-1}
\end{document}